\documentclass[a4paper,fleqn,usenatbib]{mnras}

\usepackage{graphicx}   % Including figure files
\usepackage{amsmath}    % Advanced maths commands
\usepackage{amssymb}    % Extra maths symbols
\usepackage{pdflscape}  % Landscape pages
\usepackage{subfig}
\usepackage{color}

\usepackage[T1]{fontenc}
\usepackage{ae,aecompl}
\title[Monte Carlo simulations of GC populations]{Monte Carlo simulations of multiple populations in globular clusters: constraints on the cooling flow vs. accretion scenario using million bodies simulations}
\author[Sollima]{A. Sollima$^{1}$\thanks{E-mail:
antonio.sollima@inaf.it}\\
$^{1}$ INAF Osservatorio di Astrofisica e Scienza dello spazio di Bologna, 
via Gobetti 93/3, 40129 Bologna, Italy\\
}

\date{Accepted 2021 January 15. Received 2021 January 15; in original form 2020 November 9}

\pubyear{2020}

\begin{document}
\label{firstpage}
\pagerange{\pageref{firstpage}--\pageref{lastpage}}
\maketitle

%\label{firstpage}

\begin{abstract}
I simulate the evolution of a stellar system hosting two stellar populations 
whose initial set up is defined according to the two main scenarios proposed 
for the origin of multiple populations in Galactic globular clusters: {\it (i)} formation 
of a second generation from a cooling flow of pristine+polluted gas and {\it (ii)} accretion 
of polluted gas onto the proto-stellar disks of a fraction of low-mass stars. 
For this purpose, Monte Carlo simulations containing from $10^{5}$ up to 
$3\cdot 10^{6}$ particles have been run including the effect of stellar evolution, 
binary interactions, external tidal field and a detailed modelling of the 
proto-stellar disk structure. 
The early accretion of gas onto proto-stellar disks is unable 
to produce discrete populations and to alter the 
chemical composition of a significant ($>10\%$) fraction of stars unless a disk 
lifetime larger ($t_{disk}\sim20~Myr$) than that predicted by models is assumed. 
Moreover, in this scenario the mixing timescale of the two populations is too 
short to reproduce the observed segregation of the chemically enriched population.
On the other hand, simulations run within the cooling flow scenario can evolve after a Hubble time 
into stellar systems with a first-to-second population mass 
ratio similar to that observed in globular 
clusters, provided that an initial filling-factor $r_{h}/r_{J}>0.15$ is adopted.
However, in the weak tidal field regime a radial segregation of the second population stronger 
than what observed in Milky Way globular clusters at large Galactocentric distances is predicted. 
This discrepancy disappears in simulations following eccentric orbits in a realistic axisymmetric potential.
\end{abstract}


\section{Introduction}
\label{intro_sec}

Since the beginning of the new century, the existence of multiple populations in 
Galactic globular clusters (GCs) emerged and revolutionized our view of these stellar systems \citep{gratton2019}.
In particular, the milli-mag precision of the photometers on-board the Hubble 
Space Telescope allowed to distinguish a series of discrete sequences in various regions of the 
colour-magnitude diagrams of almost all known GCs \citep[][and references therein]{piotto2006,piotto2015,milone2017}.
Stars in different sequence are characterised by different abundance of light 
elements (He, C, N, O, Mg, Al, Na) often anticorrelated between them, while iron-peak and 
$\alpha$-elements appears rather homogeneous \citep{carretta2009}.
This evidence suggests that GCs underwent a process of self-enrichment where 
new stellar populations originated from gas enriched in p-capture elements by the 
evolution of stars of previously formed populations.

The stars displaying an evolved chemical composition (with enhanced He, Na, Mg and 
depleted O and Al, hereafter referred as "second population"; SP) are generally more 
concentrated than the stars with canonical abundance patterns ("first population"; FP) \citep{lardo2011,dalessandro2019}.

Moreover, the distribution of stars in the Na-O and Mg-Al abundance planes cannot be reproduced by a simple closed box 
model where all the material expelled by a given polluter is recycled in purity, regardless of the nature of the polluter.
This implies that the polluted gas has been diluted with a consistent fraction of pristine gas \citep{dercole2010}.

The lack of an extended Main Sequence turn-off morphology together with the observed 
abundance pattern \citep[anomalies requiring p-cycles occurring at high temperatures reachable only in massive stars]{langer1993} 
suggest a fast enrichment occurring at early stages ($t<100~Myr$) led by the evolution 
of relatively massive ($M>5~M_{\odot}$) stars \citep{renzini2008}.
However, while in a stellar population characterised by a standard initial mass 
function \citep{kroupa2001} these stars release less than 30\% of the total mass budget, the SP 
generally constitutes more than 60\% of the whole cluster population \citep{carretta2009,milone2017}. 
To overcome this problem (often referred as "the mass budget problem"), many 
hypotheses have been put forward \citep{bastian2018}, which can be schematically divided in two main scenarios:
\begin{itemize}
\item{{\it The cooling flow} scenario: after the formation of the FP, the cool gas flowing from 
the evolution of some massive polluter accumulates in the central region of the 
cluster where it is mixed with a fraction of pristine gas and forms the SP \citep{calura2019}.
The majority ($>90\%$) of the FP is then lost because of the stellar evolution-driven expansion of the system, 
leaving a dominant SP surrounded by a halo of FP stars.
The subsequent dynamical evolution, driven by two-body relaxation, tends to slowly mix the two 
populations on timescales of several Gyrs, leaving after a Hubble time a GC-like object with the 
SP more abundant and still more concentrated than the FP.
Within this category, many polluters have been proposed, including intermediate-mass Asymptotic Giant Branch stars 
\citep[AGB]{dercole2008}, Fast Rotating Massive Stars \citep[FRMS]{decressin2007,krause2013} and massive interacting binaries \citep[MIB]{demink2009}.}
\item{{\it The accretion onto proto-stellar disks} scenario: the FP forms with 
a strong mass segregation, with the most massive stars located in the central region of the cluster. 
The polluted gas released in the core by these massive stars is then accreted into the proto-stellar disks of those 
low-mass stars passing through the core. Only stars with low orbital energies accrete gas and transform into SP stars, 
being therefore naturally settled preferentially in the central cluster region. 
This scenario, originally proposed by \citet{bastian2013}, has the advantage to require only the small 
amount of polluted gas which adds up to that already available in low-mass proto-stars 
which naturally constitute a reservoir of pristine gas.
Useful polluters for this scenario can be FRMS or MIB.}
\end{itemize}

In recent years, many groups attempted to test the above scenarios using chemical evolution models 
to reproduce the observed abundance patterns \citep{lind2011,dercole2012,cassisi2014}.
Unfortunately, none of the proposed polluters is able to adequately reproduce 
the distribution of stars in the Na-O and Mg-Al anticorrelations planes, unless an ad-hoc 
modification of the stellar yields is made \citep{bastian2015}.

From the dynamical point of view, the {\it cooling flow} scenario has been widely tested using 
N-body simulations focused on the derivation of the dynamical mixing timescale \citep{decressin2008,vesperini2013}, 
on the comparison of the mass function \citep{vesperini2018}, degree of rotation \citep{tiongco2019}, 
binary fraction \citep{vesperini2011,hong2015,hong2016} and relative frequency \citep{dercole2008,khalaj2015} of FP/SP stars.
All these studies are based on simulations run assuming a simplified tidal field and with 
an initial number of particles $\sim 10^{5}$, significantly smaller than the expected number of objects in real proto-GCs.
On the other hand, the predictions of the {\it accretion onto proto-stellar disks} scenario 
have been deduced only through analytical considerations in \citet{bastian2013} and \citet{henault2015}.
In both cases, low-mass stars passing through the core in an isochrone 
potential were picked as SP regardless of the amount of gas that can be 
accreted by their proto-stellar disks during the passage and without a detailed 
modelling of the proto-stellar disk structure.

All the above studies concluded that, under suitable initial conditions, both scenarios can 
reproduce the present-day general structural parameters (mass, half-mass radius) 
of GCs as well as the main observational evidence regarding the properties (relative fraction 
and radial segregation) of FP/SP, at least in a qualitative way.

In their analysis, \citet{henault2015} run N-body simulations starting from initial conditions for the FP and SP specifically 
set to mimic the predictions of both scenarios. 
From their study they found that the two scenarios predict slightly different 
residual rotation. However, the N-body simulations by \citet{tiongco2019} showed 
that, at least for the {\it cooling flow} scenario, the final rotation pattern 
of the two populations can vary according to the amount of mass loss and the 
dynamical age of the system \citep[see also][]{bellini2015}. Similarly, cluster-to-cluster variation in the differential rotation pattern of multiple populations can be naturally produced by other proposed scenarios \citep[e.g.][]{gieles2018}.
The observational evidence collected so far for 
differential rotation have indeed shown different behaviours in various GCs 
\citep{pancino2007,cordero2017,cordoni2020,bellini2018}.

One of the main difficulties in this context is due to the early epoch during 
which the formation of multiple populations took place. 
The process of formation of multiple population occurs indeed in a short timescale ($<100~Myr$) during 
which radiation, gas, and stars closely interact through complex processes like 
e.g. SNe feedback, gas-expulsion, stellar evolution-driven expansion, 
gas cooling and mixing, competitive accretion, etc.
Such a short period is then
followed by a long period during which two-body relaxation tends to erase the 
structural/dynamical differences left by the formation process.
All these processes occur with efficiencies and on timescales depending on 
different powers of the mass and radius of the stellar system, so that a 
consistent picture of the dynamical evolution of GCs can be obtained only with 
simulations with characteristics (number of particles, fraction of binaries, 
mass function) as close as possible to those of real GCs.
However, proto-GCs in the early Universe could have been as massive as $10^{7}~M_{\odot}$, orders of 
magnitudes larger than the present-day capabilities of N-body simulations.

A valuable alternative is provided by Monte Carlo simulations \citep{henon1971}. 
In this approach, the integrals of motion of particles are 
perturbed and updated at odds with N-body simulations where the orbits of individual particles need to be computed. 
This allows to use a relatively large integration time-step which fastens the computing 
time by more than one order of magnitude. 
At the present day, the largest Monte Carlo simulations contain a few $10^{6}$ particles 
\citep{kremer2020} allowing for the first time to simulate the evolution of GC-like 
objects from their formation till the present-day.
Still, million-bodies Monte Carlo simulations of star clusters accounting for 
the formation scenarios of multiple populations are not available yet.

In this paper, I present Monte Carlo simulations of GC-like stellar 
systems whose initial conditions are set according to the two main scenarios 
proposed for the formation of multiple populations with the aim of 
investigating if they can evolve toward a structural/kinematic configuration 
consistent with the available observational evidence. 

In Sect. \ref{code_sec} the Monte Carlo code adopted for the simulations is presented. 
In Sect. \ref{setup_sec} the initial setup of simulations in the framework of the 
{\it cooling flow} and {\it accretion onto proto-stellar disks} scenarios are described.
Sect. \ref{res_sec} is devoted to the presentation of the 
results and the comparison with observational data.
I summarize the results of the paper in Sect. \ref{concl_sec}.

\section{Monte Carlo code}
\label{code_sec}

The simulations adopted in this paper are run using a modification of the Monte 
Carlo code described in \citet{sollima2014} and \citet{sollima2019}. 
The code is based on the method originally developed by \citet{henon1971} and 
refined by several authors \citep{giersz1998,joshi2000,vasiliev2015}.

Schematically, the cluster is represented by a sample of stars characterized 
by their integrals of motion (energy and angular momentum) and their mass.
Starting from an initial configuration, at each time-step the following steps are performed sequentially:
\begin{itemize}
\item{The potential is evaluated according to the masses and positions of the stars;}
\item{Stars are randomly distributed along their orbit according to their energy and angular momentum;}
\item{Stars with a contiguous ranking in distance to the centre are assumed to interact 
and the perturbations in their E and L are applied.}
\end{itemize}  
The above steps are iteratively repeated until the end of the simulation.
The time-step is chosen to ensure small perturbations at any distance from 
the centre i.e. it must be smaller than the local relaxation-time.  
A detailed description of the above technique can be found in \citet{henon1971}.

Although this speed has the cost of 
losing resolution on those processes occurring on very short timescales and 
limit the treatment of the three-dimensional structure of the cluster to 
spherical symmetry, this technique has proven to accurately follow the 
dynamical evolution depicted by N-body simulations at low-N in terms of both 
structural evolution and kinematics \citep{giersz2013}.

During the last decades, several authors added additional levels of complexity 
to this method like the presence of a mass spectrum \citep{giersz2001,joshi2001},
 the effects of stellar evolution \citep{giersz2001,chatterjee2010}, 
direct integration of binary-single and binary-binary interactions \citep{fregeau2007} and interaction with a tidal field \citep{sollima2014}.

A detailed description of the basic version of the code adopted here can be 
found in \citet{sollima2014}, where the treatment of interactions among 
single mass particles and the interaction with different kinds of 
tidal fields are described.
In a subsequent version of the code \citep{sollima2019}, the presence of a mass spectrum and the 
direct integration of single-single, binary-single and binary-binary interactions have been accounted for.

In the version of the code adopted here, two further improvements have been developed:
\begin{itemize}
\item{{\bf Stellar evolution}: the evolutionary time has been extracted from the set of 
evolutionary tracks of \citet{pietrinferni2006} with Z=0.001 and [$\alpha$/Fe]=+0.4.
After this timescale the mass of the star is reduced according to the relations of \citet{kruijssen2009}.
A kick extracted from a Maxwellian distribution and randomly oriented has been added to 
the remnant of massive ($M>8~M_{\odot}$) stars. 
The 1D dispersion of the kick velocity has been assumed to be $\sigma_{kick}=100~km/s$ 
for neutron stars (originating from the evolution of stars with 
$8<M/M_{\odot}<30$) and $\sigma_{kick}=80~km/s$ for black holes (from stars 
with $M>30~M_{\odot}$). When an evolving star is part of a binary system, the 
mass-loss and the kick are applied, the orbit of the system is followed for 
10 orbital periods and, if not broken, its new equilibrium characteristics 
(semi-major axis and eccentricity) are updated.}
\item{{\bf Three-bodies encounters}: in the densest cluster region it is 
possible that three objects simultaneously cross the same small volume for a short 
interval of time. It can happen that two objects transfer kinetic energy 
to the third, emerging as a bound system. 
This process, while generally not efficient, can become effective in dense regions and when a large 
mass contrast among the involved stars is present and represents one of the 
main channels of binary formation involving massive remnants \citep{giersz1998,morscher2013}.

At each iteration, for a given pair of neighbour particles the closest object is chosen. 
The 3D velocities of the three objects have been chosen according to their values of 
E and L and randomly oriented. The probability that, during the time-step $\Delta t$, 
the three objects cross a spherical volume of radius $r_{m}$ in the same 
crossing time interval is 
\begin{equation*}
P=\pi^{2} \sqrt{\frac{2~r_{m}^{11}}{G(m_{1}+m_{2})}} n^{2} w_{12} w_{3} (1+\beta_{12})(1+\beta_{3})\Delta t
\end{equation*} 
where $m_{1},~m_{2}$ and $m_{3}$ are the masses of the three stars, $n$ is the number density $w_{12}$ is the relative velocity of the first two stars and $w_{3}$ 
that of the third star with respect to the center of mass of the first two and
\begin{eqnarray*}
\beta_{12}&=&\frac{2 G (m_{1}+m_{2})}{r_{m} w_{12}^{2}}\nonumber\\
\beta_{3}&=&\frac{2 G (m_{1}+m_{2}+m_{3})}{r_{m} w_{3}^{2}}\nonumber
\end{eqnarray*}
I adopted a maximum impact parameter 
\begin{equation*}
r_{m}=\frac{G (m_{1}+m_{2})}{E_{1}+E_{2}-\Phi_{1}-\Phi_{2}}
\end{equation*}
where $E_{i}$ and $\Phi_{i}$ are the energy of the considered star and the 
potential at its position.
A random number uniformly distributed between 0 and 1 is extracted and, if smaller than
the associated probability, the close interaction is integrated using a symplectic 
integrator with adaptive timestep \citep{yoshida1990}.
For this purpose, the impact parameters $b_{i}$ (i=12,3) have been chosen as 
\begin{equation*}
b_{i}=r_{m}\frac{\beta_{i}}{2}\left[\sqrt{1+4\eta_{i}\left(\frac{\beta_{i}+1}{\beta_{i}^{2}}\right)}-1\right]
\end{equation*}
 where $\eta_{i}$ are random numbers uniformly distributed between 0 and 1.
The initial positions of the three stars are defined such that they 
simultaneously reach the minimum distance corresponding to the extracted impact parameters.
The interaction is followed until the total energy of the system exceeds zero. 
The velocities of the emerging objects are then converted in the cluster reference 
frame and their corresponding values of E and L are assigned accordingly.
}
\end{itemize}

\section{Initial setup}
\label{setup_sec}

In this section, the initial setup of the Monte Carlo simulations presented
in this paper is described.
Of course, some of the initial conditions vary according to the adopted 
scenario and are discussed in the following subsections. 

Unless specified otherwise, in both scenarios simulations were run with an initial FP 
mass of $M_{F}=10^{6}~M_{\odot}$, a fraction of binaries $f_{b}=5\%$, and 
a \citet{kroupa2001} Initial Mass Function (IMF) defined between 0.1 and 120 $M_{\odot}$.
The characteristics of binary stars (semi-major axis, eccentricity, mass of the components)
were chosen using the technique descirbed in \citet{sollima2012}. In particular, 
the masses of the components are chosen by random pairing stars from a 
\citet{kroupa2001} IMF, and the period and eccentricity have been extracted from the 
distribution of \citet{duquennoy1991}. From an initial library, stars with semi-major axis smaller 
than the stable overflow criterion by \citet{lee1988} or larger than the ionization limit of \citet{hut1983} are removed.
In this configuration, the total number of particles at the beginning of the simulation is $\sim 1.5\cdot 10^{6}$.
To investigate the effect of the initial mass on the long-term evolution of the two populations, 
a few simulations were run with a lower ($10^{5}~M_{\odot}$) or higher ($2\cdot 10^{6}~M_{\odot}$) initial 
mass.

\begin{figure}
 \includegraphics[width=8.6cm]{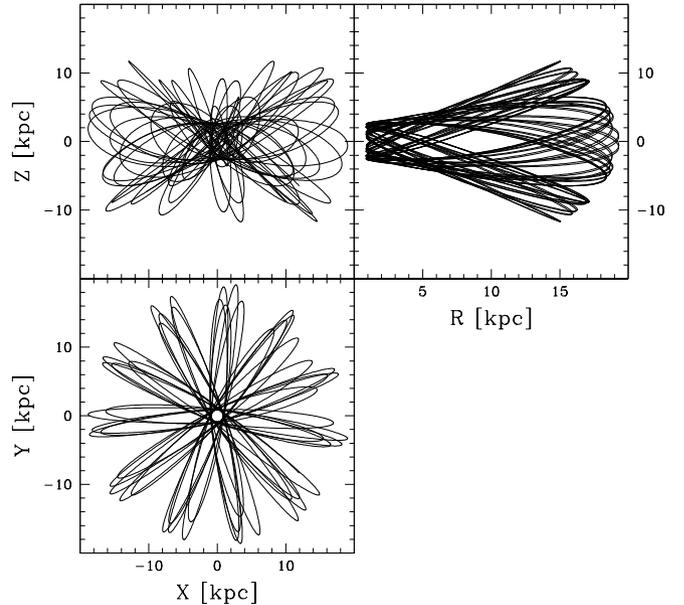}
 \caption{Adopted NGC1851-like orbit in the X-Y (left panel), X-Z (middle panel) 
 and R-Z (right panel) planes.}
\label{orb1851}
\end{figure}

The simulated clusters move in the tidal field generated by a point mass with 
$M_{g}=v_{circ}^{2}R_{g}/G$, where $v_{circ}=220~km/s$, and follow 
a circular orbit at galactocentric distance $R_{g}=14.4~kpc$.
Such a distance has been chosen to mimic the average tidal field felt by 
Galactic GCs. For this purpose, the orbits of Galactic GCs have been integrated 
within the Galactic potential of \citet{johnston1995} 
using the 3D positions and velocities listed by \citet{baumgardt2019}, and the orbit-averaged Jacobi 
radius at the conventional mass 
of $M=10^{5}~M_{\odot}$ ($\langle r_{J5}\rangle$) has been calculated by 
averaging along the orbit the instantaneous Jacobi radius ($r_{J}$) calculated 
using eq. A2 of \citet{allen2006}. Galactic GCs have values of 
$60<\langle r_{J5}\rangle/pc<120$ with an average of 
$\overline{\langle r_{J5}\rangle}=85~pc$. The same value of 
$\langle r_{J5}\rangle$ can be obtained in the simplified point-mass potential 
assuming $R_{g}=14.4~kpc$.
Of course, this crude approximation provides a guess value of $R_{g}$ to mimic the tidal 
cut imposed by the Galactic tidal field but cannot account for the tidal shocks
occurring in a realistic Galactic potential as a result of disk crossing and 
peri-Galactic passages \citep{ostriker1972,aguilar1988}.
To check the effect of the adopted tidal field, some simulations have been run adopting 
a smaller Galactocentric distance ($R_{g}=5.1~kpc$) or the tidal field generated by
the axisymmetric potential of \citet{johnston1995} and the orbit of the GC NGC1851 
(see Fig. \ref{orb1851}). This last orbit, although having a relatively large orbit-averaged 
Jacobi radius $\langle r_{J5}\rangle=109~pc$, is extremely eccentric ($e=0.9$) with a 
peri-Galactic distance of $\sim$1 kpc and an orbital period of $\sim$450 Myr, 
so that some 60 bulge- and 120 disk- shocking occur during the entire cluster evolution.

The simulations are run for 12 Gyr allowing to follow the evolution of the 
relative fraction and of the various structural properties of FP and SP.

The initial and final properties of the entire set of simulations are listed in 
Table \ref{table1_tab}.

\subsection{Cooling flow scenario}
\label{cool_setup_sec}

The simulations have been preformed in the framework of the scenario proposed 
by \citet{dercole2008}. In this model, the self-enrichment occurs in the central 
region of the cluster where the gas expelled by intermediate-mass AGB stars cools and mix 
with a fraction of pristine gas, residual from the formation of the FP. 
According to \citet{dercole2010}, assuming ad-hoc yields for AGB and super-AGB stars, 
it is possible to reproduce the distribution of stars in the Na-O anticorrelation plane 
assuming a significant contribution of pristine gas such that the emerging SP has a mass 
$\sim$10\% of that of the FP. The formation of the SP occurs after $\sim$30 Myr when 
massive AGB stars start to evolve and ends after $\sim$70 Myr.
Of course, if different polluters are considered \citep[like e.g. the FRMS;][]{krause2013}, 
the onset and the duration of the SP formation burst change. However, since the entire star 
formation process is expected to end after a $<100$ Myr in all the considered scenarios, 
the detailed timing of this process should not have a significant effect on the final outcome 
of the simulation after a Hubble time.

According to the above picture, FP stars were extracted from the adopted IMF and 
distributed according an isotropic \citet{king1966} \citep[][with $W_{0}=25$ for simulations assuming primordial mass segregation]{gunn1979} 
density profile. In one case a FP with a large degree of radial anisotropy \citep[$r_{a}/r_{h,F}=1$; where $r_{a}$ is the anisotropy 
radius beyond 
which stellar orbits progressively start to be radially biased;][]{gunn1979} has been 
simulated.
The half-mass radius of the FP has been chosen to ensure the large initial Roche-lobe 
filling factor ($0.1<r_{h}/r_{J}<0.2$) needed to ensure a quick loss of FP stars\footnote{For reference, a \citet{king1966} model with $W_{0}=5$ and $r_{h}/r_{J}\sim0.2$ 
has a tidal radius equal to the Jacobi radius.}. 

Between 30 Myr and 100 Myr after the beginning of the simulation, SP stars are 
continuously added in the central region of the 
cluster following a \citet{king1966} profile with central adimensional 
potential $W_{0}=5$ and a half-mass radius corresponding to 10\% of that of 
the FP ($r_{h,S}=0.1~r_{h,F}$ in all simulations but two where different value of 
$r_{h,S}/r_{h,F}=0.05$ and 0.2 are explored). The energies and angular momenta of SP stars are drawn solving the anisotropic Jeans equation 
using the instantaneous cluster potential and adopting an Osipkov-Merrit radial 
anisotropy profile \citep{osipkov1979,merritt1985} with anisotropy radius equal to $r_{a}=r_{h,S}$. 
In one case the degree of anisotropy of FP and SP have been inverted, with a radially anisotropic FP 
and an isotropic SP (see above).
The SP is constituted by stars extracted from a \citet{kroupa2001} IMF between 0.1 and 8 $M_{\odot}$ \citep[to avoid the explosion of SP SNe II;][]{dercole2008}
and contains the same fraction of binaries with the same characteristics of those in the FP. 
The rate at which SP stars are added is $dM/dt=0.00143~M_{F}~Myr^{-1}$, so that at the 
end of the SP formation episode (after 100 Myr from the beginning of the simulation) 
$M_{S}=0.1~M_{F}$.
Note that, given the large uncertainties on the AGB and super-AGB yields \citep{ventura2013,bastian2015}, as well as the poor 
knowledge of the origin of the dilution process, the constraint on the initial fraction of SP/FP is extremely weak, 
and other fractions have been proposed \cite[e.g.][]{cabrera2015}. 
Of course, a smaller (larger) initial SP/FP fraction would require a more (less) 
efficient loss of FP stars i.e. a larger (smaller) initial Roche-lobe filling factor.
Given the degeneracy between these two parameters and considering the uncertainties 
in the details of the dilution process, it is impossible to place any constraint on the initial SP/FP 
fraction. For this reason, I kept fixed this parameter.

\subsection{Accretion onto proto-stellar disks scenario}
\label{disk_setup_sec}

\begin{figure}
 \includegraphics[width=8.6cm]{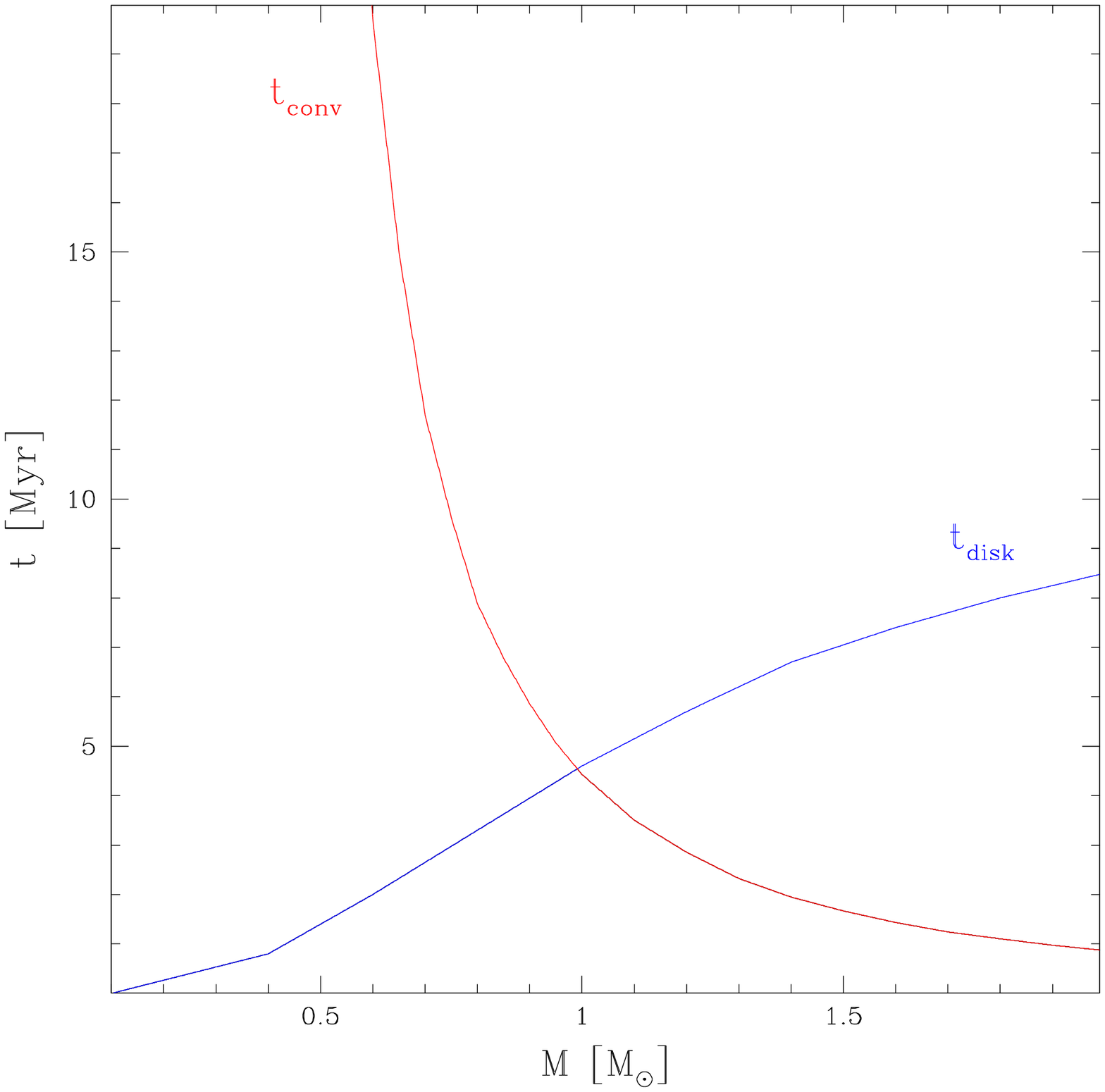}
 \caption{Proto-stellar disk lifetime 
 \citep[$t_{disk}$; from][blue line]{li2016} and duration of the fully 
 convective phase \citep[$t_{conv}$; from][red line]{tognelli2011} as 
 a function of the stellar mass.}
\label{times}
\end{figure}

\begin{figure*}
 \includegraphics[width=\textwidth]{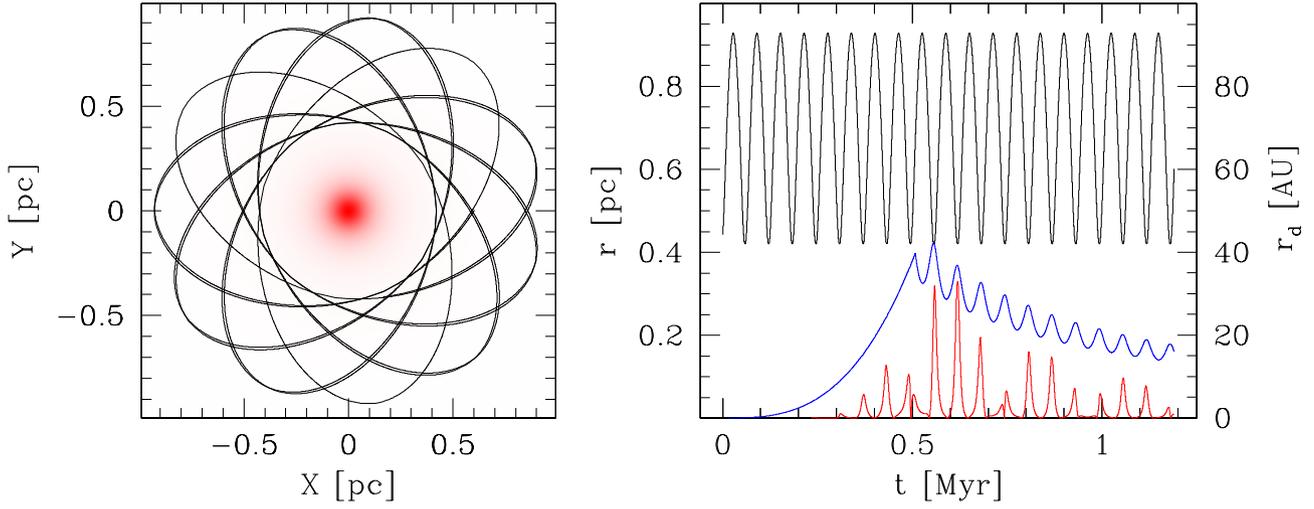}
 \caption{Orbit of a sample star in simulation R14.4lm6rh1 (left panel). 
 The location of the gas is marked by the red area. The corresponding evolution of 
 the distance from the cluster centre (black), 
 the proto-stellar disk truncation radius (blue) and the accretion rate (red) 
 are shown in the right panel.}
\label{mibfig}
\end{figure*}

\begin{figure}
 \includegraphics[width=8.6cm]{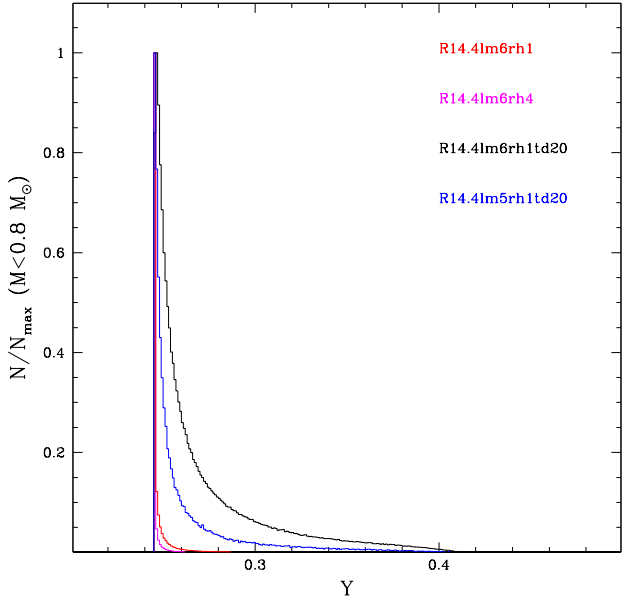}
 \caption{Distribution of He mass fration in the simulations run within the 
 {\it accretion onto proto-stellar disks} scenario. 
 All histograms are normalized to the peak value.}
\label{mibhis}
\end{figure}

\begin{figure*}
 \includegraphics[width=\textwidth]{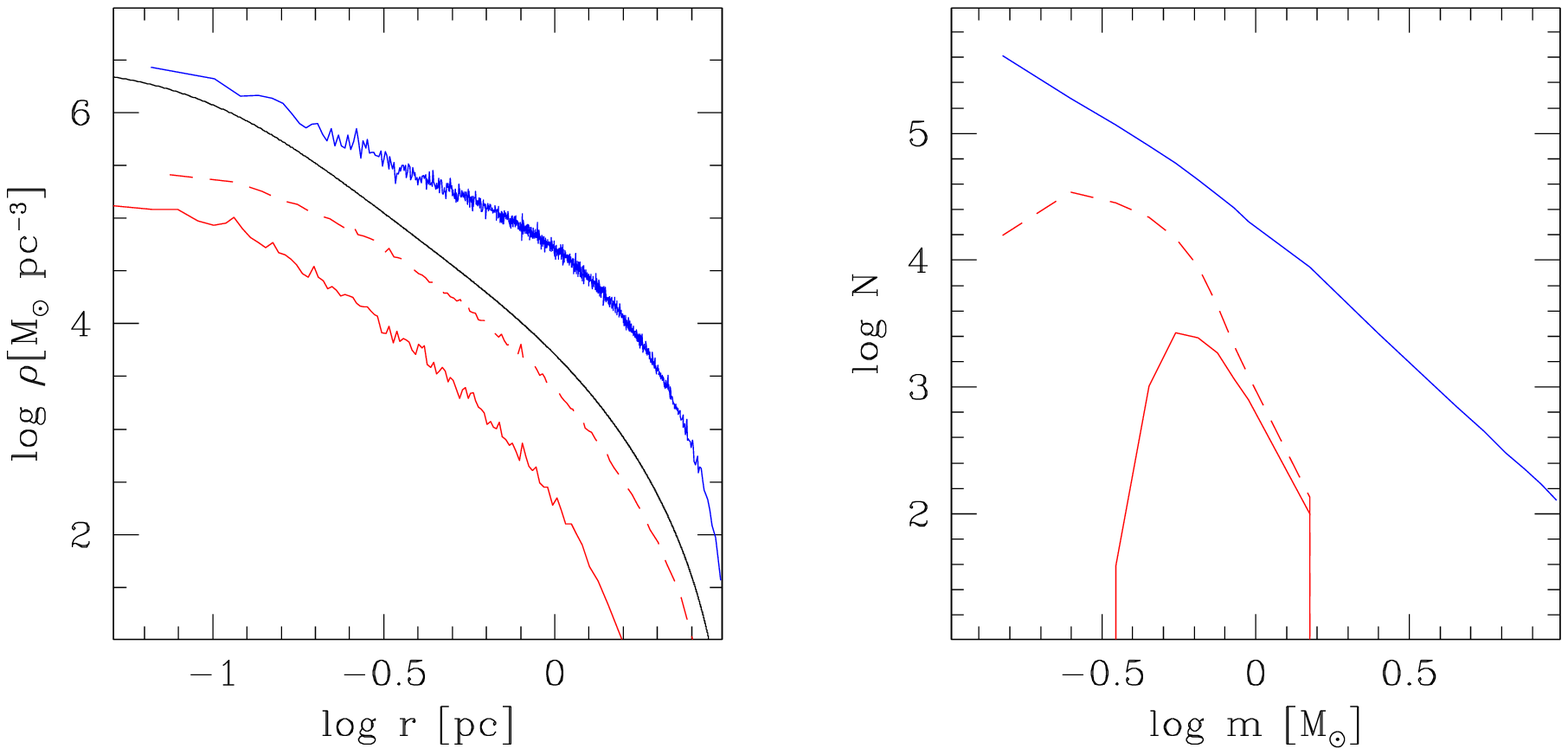}
 \caption{Initial density (left panel) profiles and mass functions (right panel) of FP (blue lines) and 
 SP (red lines) in the R14.4lm6rh1 (solid lines) and R14.4lm6rh1td20 (dashed lines) 
 simulations. The gas density profile is overplotted in the left panel with a 
 black line.}
\label{mibprof}
\end{figure*}

In the scenario proposed by \citet{bastian2013}, there is only one star formation 
episode where the FP forms in a mass segregated configuration. 
After a few Myr massive ($>10~M_{\odot}$) stars evolve and release part of their gas in the central 
region of the cluster. Low-mass stars, still in their pre-Main Sequence phase, 
cross this region and accrete this gas onto their proto-stellar disks, mixing it with their own gas. 
To account for the abundance variation of elements involved in the nuclear burning (e.g. He), it is 
necessary that the accreted material is brought from the surface to the stellar 
core, an occurrence possible only until the star is entirely convective \citep{dantona2014}.

To calculate the amount of gas accreted by low-mass stars, I performed an analytical 
modeling of the disk structure.

The initial population is distributed in 
the phase-space according to a multi-mass models of \citet{gunn1979} with $W_{0}=25$.
For this purpose, the simulated stellar population has been grouped in 29 mass bins, 
where stars with masses $M>10~M_{\odot}$ are in the last 11 bins.

The gas cloud density profile has been derived as 
\begin{equation*}
\rho_{g}=\sum_{i=19}^{29}\mu_{i} n_{i} m_{i}
\end{equation*}
where $m_{i}$ is the mass, $\mu_{i}$ is the fraction of released gas \citep[from][]{kruijssen2009}, 
$n_{i}$ is the number density of the stars in the i-th mass bin.

The orbits of less-massive ($M<10~M_{\odot}$) stars within the cluster potential are followed using 
a 4th-order Runge-Kutta integrator for the time their proto-stellar disks are able 
to accrete gas ($t_{end}$).
This timescale has been calculated as the minimum between the disk lifetime 
($t_{disk}$) and the time during which the star is entirely convective ($t_{conv}$).
I adopted the disk lifetimes calculated by \citet{li2016} (LX16) as a function of the 
stellar mass. According to these authors, the proto-stellar disk survives 
longer in massive stars than in low-mass ones, ranging from 0.8 Myr in a 0.4 $M_{\odot}$ 
star to 8.5 Myr in a 2 $M_{\odot}$ star.
Note that these models are built assuming the disk formed by gas in stable 
rotation. This condition can be altered in the case of a strong accretion of 
gas at low angular momentum, so that the stability of the disk is not 
guaranteed \citep{wijnen2016}.
The time a star maintain a fully convective structure has been taken from the pre-Main 
Sequence evolutionary tracks of \citet{tognelli2011} with a metal content Z=0.001.
The dependence of $t_{disk}$ and $t_{conv}$ as a function of the stellar 
mass is shown in Fig. \ref{times}. It can be seen that the time during which the 
proto-star is able to accrete gas and use it as a fuel in its core is a peaked 
function of mass with a maximum duration of $\sim5$ Myr at the mass of 
$M\sim 1~M_{\odot}$. As hypothesized by \citet{bastian2013}, the gas accretion can prolong the 
disk lifetime. To account for this effect, the orbit integration is interrupted when the 
integration time exceeds the instantaneous value of $t_{end}(M(t))$.
Moreover, to account for the uncertain modeling of disk lifetimes, additional simulations 
were performed assuming an increased lifetime $t_{disk}=20~Myr$ regardless of the stellar mass.

The gas crossing the disk during this time interval (i.e. the maximum amount of 
gas that can be accreted by the star) is given by
\begin{equation*}
\Delta M=\int_{0}^{t_{end}} \pi~r_{d}^{2}~\rho_{g}~v~\cos{\alpha}~dt
\end{equation*}
where $v$ is the modulus of the star velocity, $r_{d}$ is the proto-stellar 
disk radius and $\alpha$ is the angle between the star position and velocity 
vectors. All the above quantities are time-dependent and are calculated 
along the star orbit.

In case of isolated stars, the models of \citet{nakamoto1994} and \citet{li2016} indicate that the
disk radius grows with time until a gravitational instability occurs or 
X-ray photoevaporation destroy the outer regions 
of the disk \citep{owen2012}, at distances as large as 50 AU. 
This is confirmed by the existence of extremely extended disks 
\citep[up to 1000 AU;][]{lada2000} around young stars in the Trapezium cluster. 
However, within a dense cluster the main truncation mechanism for disks is the 
close passage of nearby stars. To account for this effect I applied a Monte 
Carlo scheme to implement the technique developed by \citet{dejuan2012} locally 
along the orbit followed by the star inside the cluster.
At each point along the orbit, $10^{5}$ synthetic particles have been 
extracted with the local mass function and the corresponding velocity distribution.
An impact parameter 
\begin{equation*}
b=b_{max}\sqrt{\eta}
\end{equation*}
has been extracted, with $\eta$ being a random number uniformly 
distributed between 0 and 1, $n$ is the local number density and 
\begin{equation*}
b_{max}=\left(\frac{48}{\pi~n}\right)^{1/3}
\end{equation*}
The truncation radius implied by the encounter with a star colliding with the extracted mass $m'$, 
impact parameter $b$, and a relative velocity with modulus $w$, is given by
\begin{equation*}
r_{tr}'(w,m',b)=\frac{G (m+m')}{w^{2}(1+\sqrt{m'/m})}\left(\sqrt{1+\frac{b~w^{2}}{G (m+m')}}-1\right)
\end{equation*}
The total number of collisions occurring with impact parameter $b<b_{max}$ before 
the time $t$ is instead given by
\begin{equation*}
N_{coll}=0.3 \pi b_{max}^{2} n \langle w\rangle t
\end{equation*}
where $\langle w\rangle$ is averaged over the $10^{5}$ synthetic particles and 
the coefficient 0.3 is a corrective factor accounting for the mild effect of 
encounters occurring with inclination angles $>45^{\circ}$ \citep{dejuan2012}.
The individual values of $r_{tr}'$ are sorted and a relation between the 
ranking index and $r_{tr}'$ is derived.
The instantaneous truncation radius ($t_{tr,t}$), set by the most likely 
closest collision, is then defined by interpolating 
through the above relation at the index $i'=10^{5}/N_{coll}$.

The disk radius at the given time and in a given point along the star orbit is finally given by
\begin{equation*}
r_{d}=min\left[31 \left(\frac{t}{5\cdot 10^{5}yr}\right)^{3} pc,~r_{tr,t}\right]
\end{equation*}

To compute the actual enhancement provided by the accretion process, the total He mass fraction 
has been calculated assuming a cosmological He abundance for the FP \citep[Y=0.245;][]{planck2016} 
and the extreme He abundance predicted by \citet{demink2009} 
\begin{equation*}
Y=\frac{0.245~M_{0}+0.44~\Delta M}{M_{0}+\Delta M}
\end{equation*}
where $M_{0}$ is the original stellar mass before accretion.

The evolution of the accretion rate of a sample star in the R14.4lm6rh1td20 
simulation during its first Myr is shown in Fig. \ref{mibfig}.
This star, after moving for 20 Myr within the inner pc, reaches a final mass 
of 0.14 $M_{\odot}$ and a He mass fraction $Y=0.3$.
It can be noted that the maximum accretion occurs during pericenters, where the 
gas density and the star speed are high. The accretion rate is however 
modulated according to the size of the proto-stellar disk which increases at the beginning of the disk evolution, 
reaches a maximum and then progressively declines as a result of the cumulative 
effect of close collisions. Consequently, most of the accretion occurs at early epochs, 
while after a few Myr the disk is almost unable to accrete gas anymore. 

The distributions of He mass fractions among low-mass ($M<0.8~M_{\odot}$; i.e. those 
expected to survive after a Hubble time) for the four considered initial 
conditions are shown in Fig. \ref{mibhis}. 
It is immediately apparent that in all simulations, the distribution is unimodal with
a peak at the cosmological value ($Y=0.245$) and a tail toward high Y 
mass fractions. 
In the absence of any apparent bimodality, I conventionally assigned stars with $Y>0.3$ to the SP.
In all simulations adopting the disk lifetimes predicted by the LX16 
model the fraction of SP stars is negligible ($M_{S}/M_{tot}<1\%$).
A slightly larger fraction of SP ($\sim 8\%$) is instead predicted by simulations adopting 
an extended disk lifetime, although it is far below the observed value in the majority 
of GCs \citep[$>60\%$;][]{carretta2009,milone2017}.
Note also that, for a fixed half-mass radius, the larger the 
initial cluster mass the larger is the abundance of SP, in agreement with what observed 
among GCs \citep{carretta2010}. This is due to the increasing density and mean 
velocity in massive clusters, leading to large accretion rates.
For the same reason, the simulation starting with a large half-mass radius (R14.4lm6rh4) has a 
smaller fraction of SP stars with respect to the simulation with a small 
half-mass radius and the same mass (R14.4lm6rh1).

The initial density profiles of gas, FP and SP for the simulation R14.4lm6rh1 
and R14.4lm6rh1td20 are shown in the left panel of Fig. \ref{mibprof}. 
It can be seen that, in both simulations, the SP naturally forms more 
concentrated than FP, and follows a density profile similar to that of the polluted gas. 
However, only a small fraction of the gas is used to form the SP.
Another peculiarity of this scenario is related to the resulting mass functions of
the two populations (see the right panel of Fig. \ref{mibprof}). Indeed, while the FP
forms with the standard \citet{kroupa2001} IMF, the SP is able to accrete efficiently 
gas only in a restricted range of mass, thus following a mass function peaked at 
$log~(M/M_{\odot}\sim-0.3$, 
with a steep slope $\alpha\sim-5$ at masses $-0.3<log(M/M_{\odot})<0.2$ 
and deprived of stars outside the mass range $-0.5<log(M/M_{\odot})<0.2$.
Qualitatively, the same evidence is apparent in simulations where the disk 
lifetime has been increased to 20 Myr at all masses. In this case, low-mass stars 
have more time to accrete gas, so that the SP is more abundant and the mass function 
peak is shifted at a lower mass.

\section{Results}
\label{res_sec}

\subsection{Cooling flow scenario}
\label{cool_res_sec}

\begin{figure}
 \includegraphics[width=8.6cm]{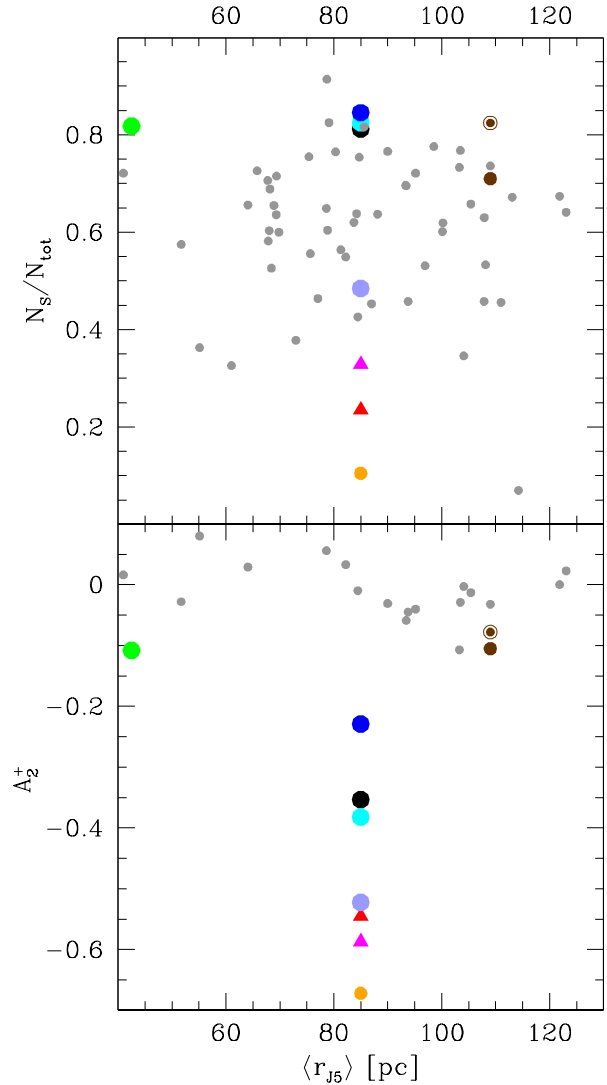}
 \caption{Fraction of SP (top panel) and $A_{2}^{+}$ parameter (bottom panel) 
 measured in the last snapshot of all the simulations run within the 
 {\it cooling flow} scenario.
 Grey points represents the observational fractions for the Galactic GCs \citep[from][]{milone2017,dalessandro2019}. 
 The colour code is described in the last column of Table \ref{table1_tab}. Circles and triangles represent 
 simulations with and without primordial mass segregation. The symbol size is proportional to the initial Roche-lobe filling factor $r_{h}/r_{J}$.}
\label{agbres}
\end{figure}

\begin{figure}
 \includegraphics[width=8.6cm]{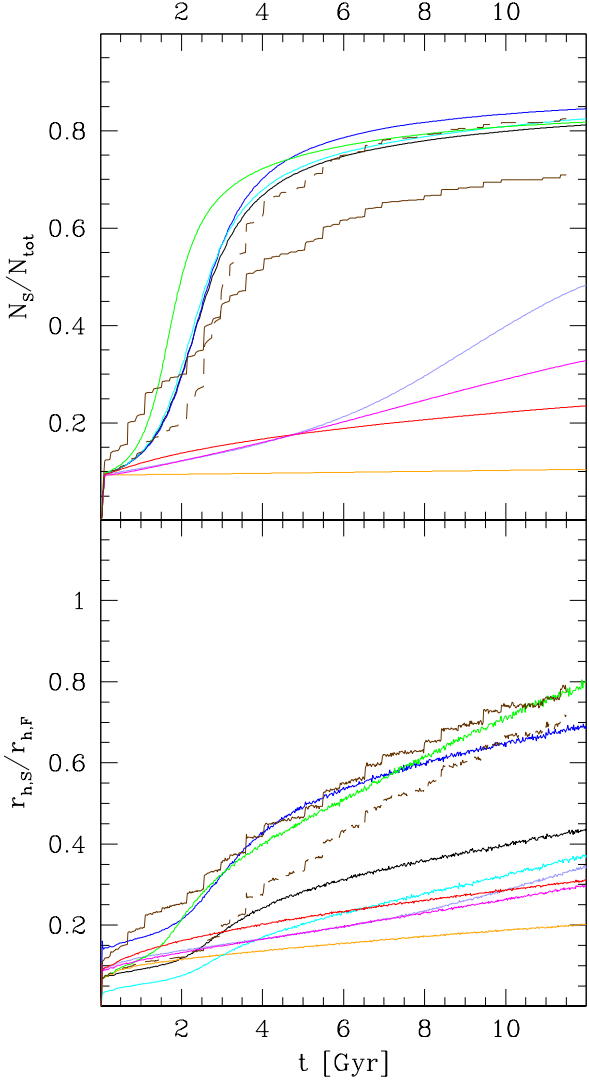}
 \caption{Evolution of the SP number fraction (top panel) and SP/FP half-mass 
 radii (bottom panel) for all the simulations run within the 
 {\it cooling flow} scenario.
 The colour code is described in the last column of Table \ref{table1_tab}.}
\label{agbev}
\end{figure}

\begin{figure}
 \includegraphics[width=8.6cm]{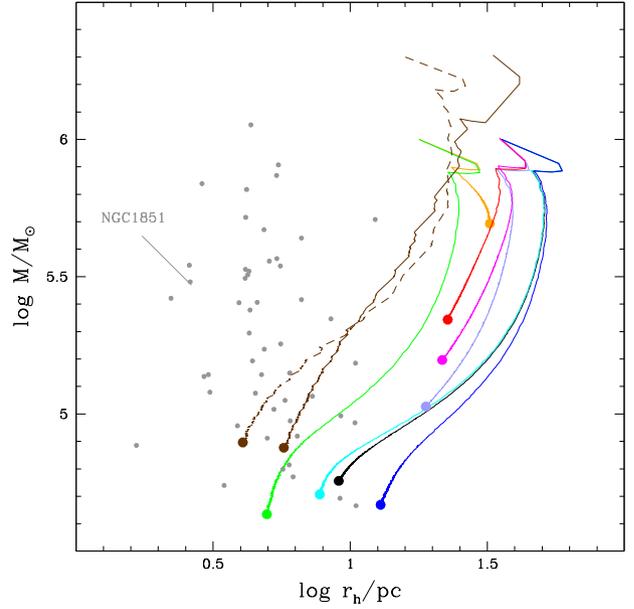}
 \caption{Evolution of all the simulations run within the {\it cooling flow} scenario
 in the $M-r_{h}$ plane. The large dots mark the endpoints of the simulations. 
 Grey points represents the Galactic GCs with $70<\langle r_{J5}\rangle/pc<100$. 
 The location of NGC1851 is also indicated.
 The colour code is described in the last column of Table \ref{table1_tab}.}
\label{agbmr}
\end{figure}

\begin{figure*}
 \includegraphics[width=\textwidth]{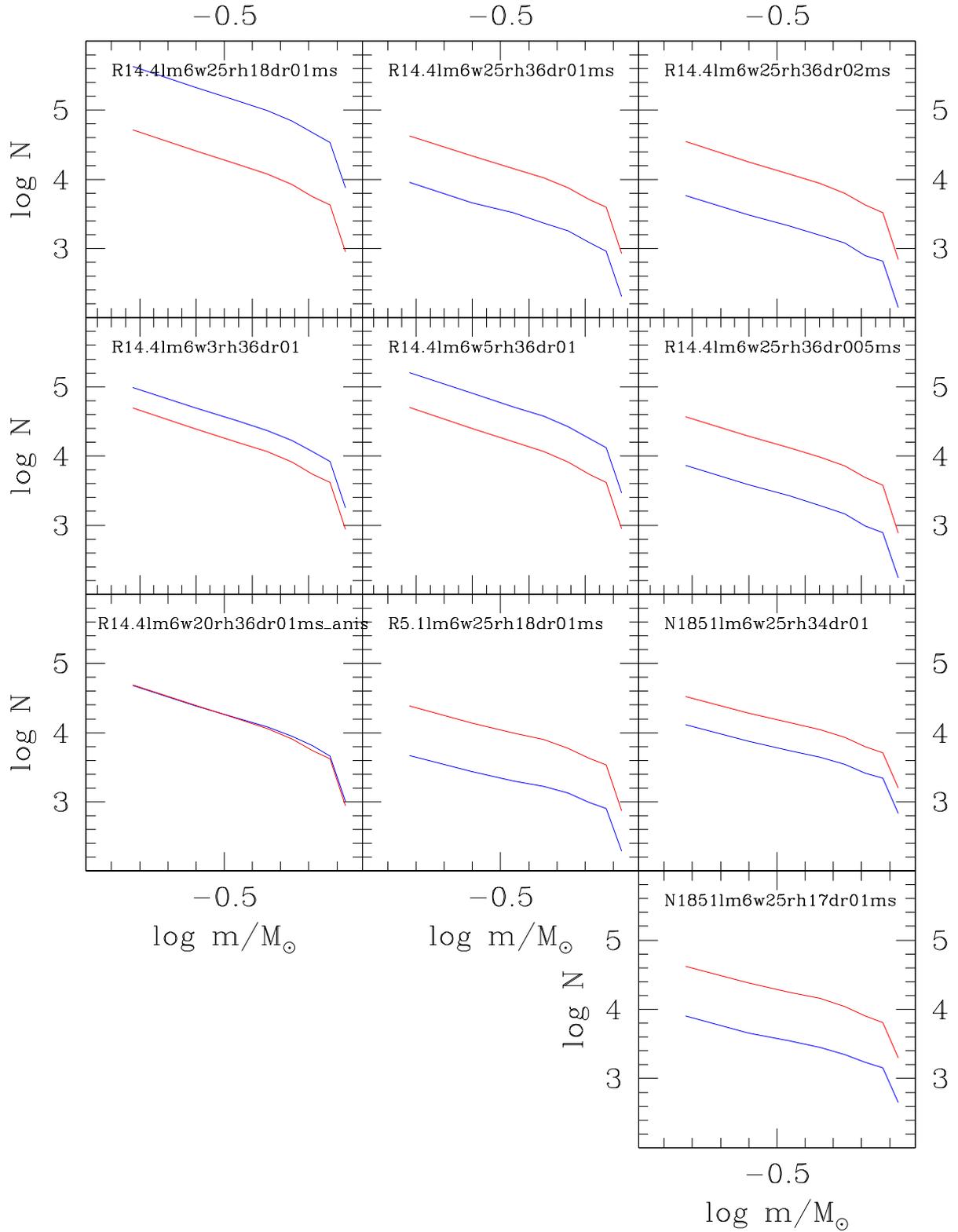}
 \caption{Mass function of FP (blue line) and SP (red line) measured in the last snapshot 
 of all the simulations run within the {\it cooling flow} scenario.}
\label{agbmfall}
\end{figure*}

\begin{figure*}
 \includegraphics[width=\textwidth]{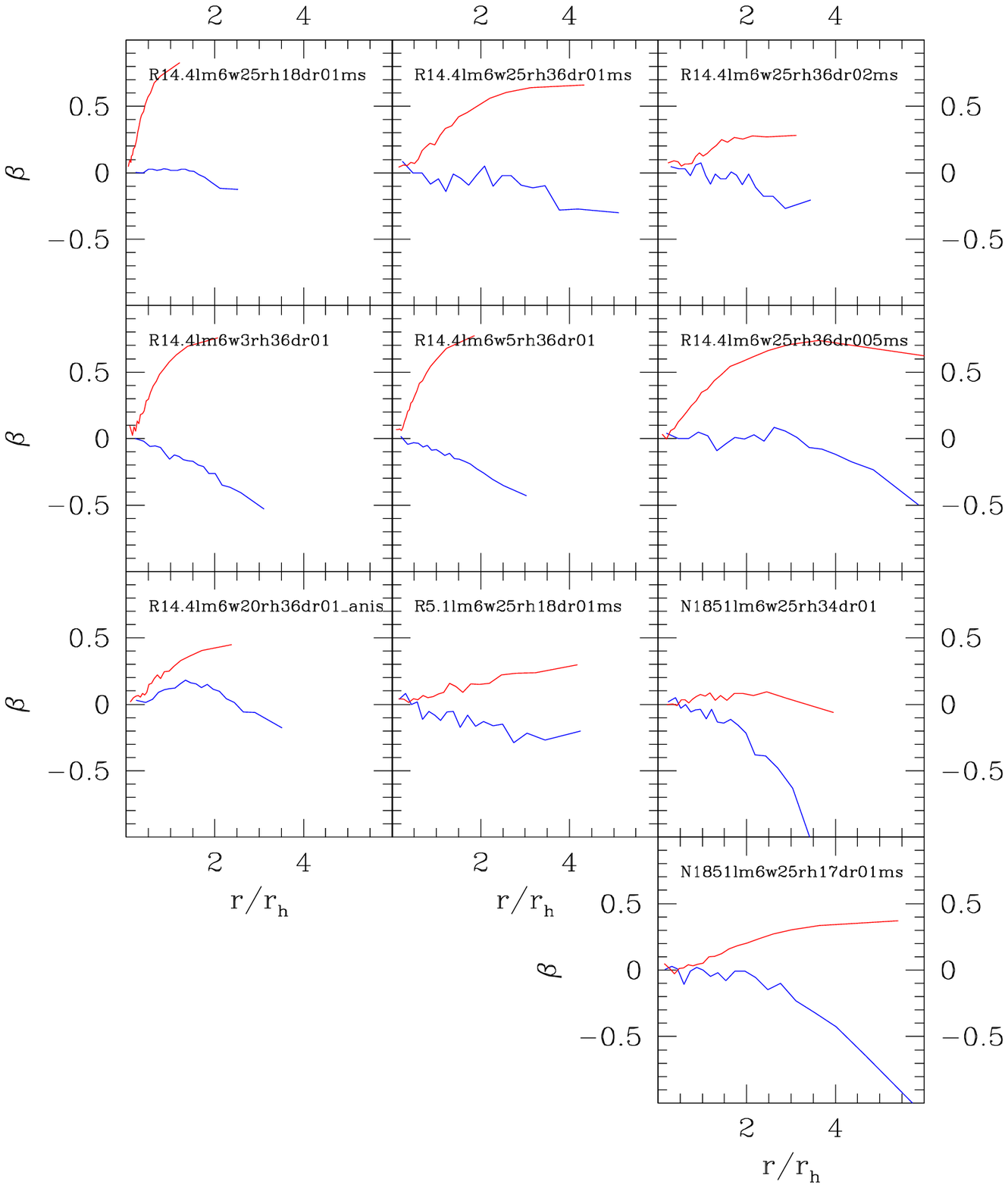}
 \caption{Anisotropy parameter of FP (blue lines) and SP (red lines) as a 
 function of the distance from the 
 cluster centre measured in the last snapshot of the simulations run
 according to the {\it cooling flow} scenario. 
 Distances are normalized to the final global half-mass radius.}
\label{agbball}
\end{figure*}

In the top panel of Fig. \ref{agbres} the fractions of SP stars in the 
last snapshot of the 10 simulations run within the {\it cooling flow} scenario are 
plotted against the
orbit-averaged Jacobi radius (i.e. indicating the strength of the tidal field 
they are subject to) and compared with the fractions measured in 
Galactic GCs by \citet{milone2017}. 
Among simulations moving on a circular orbit, it can be noted that both simulations 
starting without primordial mass segregation (R14.4lm6w5rh36dr01 and 
R14.4lm6w3rh36dr01) predict a fraction of SP stars $N_{S}/N_{tot}<0.4$, 
significantly smaller than that observed in real GCs. 
Between them, the simulation with a shallower density profile (R14.4lm6w3rh36dr01) loses slightly 
more FP stars than the one with the steeper profile (R14.4lm6w5rh36dr01). 
This is expected since in the former simulation a larger fraction of FP stars is located 
at large distances from the cluster centre with respect to the latter simulation which 
is more concentrated and therefore more resistant to evaporation \citep[see also][]{vesperini1997}.
Among simulations with primordial mass segregation, the fraction of SP stars 
presents a strong dependence on the Roche-lobe filling factor, 
with the most tidally-filling simulations ($r_{h}/r_{J}\geq 0.15$) able to reach the large 
observational values.
Both the above evidences are expected as a result of the increasing amount of 
mass loss (mainly FP stars) experienced by simulations starting with primordial mass segregation 
and with a large filling factor. Indeed, the potential energy lost in the cluster 
centre is given by $\Delta\Phi=\sum G \Delta m_{i}/r_{i}$, where $\Delta m_{i}$ is 
the mass lost by the i-th star because of stellar evolution and $r_{i}$ is the distance 
of the star from the cluster centre. In mass segregated simulations massive stars (subject to 
a large amount of mass loss) are preferentially located at small distances, so 
that the loss of potential energy is larger than in non-segregated simulations.
These simulations strongly expand to maintain the virial equilibrium pushing FP stars, 
preferentially located in the peripheral cluster region, outside the Jacobi radius.
In the same way, the larger is the initial Roche-lobe filling factor the larger 
is the fraction of stars exceeding the Jacobi radius during this early expansion.
Instead, SP stars segregated in the cluster centre remain at the bottom 
of the potential well and are more efficiently retained until the end of the simulation.
The subsequent evolution has only a minor effect in the global fraction of SP stars.
This process is depicted in the top panel of Fig. \ref{agbev}, where the evolution of 
the number ratio of FP and SP stars of the 10 simulations are compared.  

To investigate the degree of mass segregation of SP stars, the parameter 
$A_{2}^{+}$ \citep{dalessandro2019} has been calculated. 
For this purpose, the 3D positions of particles with masses $m>0.7~M_{\odot}$ have 
been projected on the plane of the sky assuming an isotropic spherical 
symmetry. The cumulative distributions of FP/SP stars located within $2~r_{h}$ from the cluster centre, 
have been computed and normalized to their respective total numbers. 
The $A_{2}^{+}$ parameter is then calculated as the difference of the integrals 
of the two distributions. In the bottom panels of Fig. \ref{agbres} the values of $A_{2}^{+}$
computed on the last snapshot of the 10 simulations are plotted against 
$\langle r_{J5}\rangle$ and compared with those estimated by 
\citet{dalessandro2019} for 18 Galactic GCs.
Note that, while real GCs show values of $-0.11<A_{2}^{+}<0.08$ with a weak decreasing 
trend with $\langle r_{J5}\rangle$, at the end of all the simulations moving on circular orbits at a distance 
$R_{g}=14.4~kpc$ the SP is much more segregated in the centre ($A_{2}^{+}<-0.2$).
The evolution of the ratio of the SP and FP half-mass radii in the various simulations 
is shown in the bottom panel of Fig. \ref{agbev}. It is apparent that SP expands at most by a factor 
of 4 at the end of these simulations. 
This is a consequence of the long relaxation time of these simulations which 
reduces the efficiency of the dynamical mixing between the two populations.
The half-mass relaxation time \citep{spitzer1987} is indeed $t_{rh}>20~Gyr$ at the beginning of all these 
simulations and remains longer than 3 Gyr during their entire evolution.
Note that the dynamical mixing efficiency of the two populations depends on the tidal truncation and not on the 
initial half-mass radius. 
Indeed, the same discrepancy is present in the underfilling simulations 
R14.4lm6w25rh18dr01ms which, in spite of its
relatively short initial half-mass relaxation time, 
quickly expand without loosing a significant amount of mass, thus increasing their 
half-mass relaxation time during their evolution.
A better agreement is instead found with simulation R5.1lm5w25rh18dr01ms 
starting with a small half-mass radius while being tidally-filling. In this last case, 
the tidal cut keeps the simulation compact with a relaxation time which is short enough to 
allow the two populations to efficiently mix.
It is also interesting to note that, comparing simulations R14.4lm6w25rh36dr005ms, R14.4lm6w25rh36dr01ms and
R14.4lm6w25rh36dr02ms, the initial concentration of the SP has only a minor 
effect on both the SP fraction and on its radial segregation. 
In fact the SP/FP ratio mainly depends on the efficiency of evaporation of the FP, 
while the SP is almost unaffected by mass loss.
Moreover the SP, when initially more concentrated, expands faster and partially 
compensates for the initial condition.  

In Fig. \ref{agbmr} the evolution of the 10 simulations in the mass vs. 
half-mass radius plane is shown. For comparison, the location 
of Galactic GCs with $70<\langle r_{J5}\rangle/pc<100$ is shown (see Sect. \ref{setup_sec}).
The evolution of all simulations follows in this plane a characteristic path: an initial 
expansion (due to the stellar evolution-driven mass loss) followed by a temporary 
contraction of the half-mass (due to the formation of the SP in the centre), and a subsequent 
phase where both mass and half-mass radius decrease. The length of the path depends mainly 
on the initial Roche-lobe filling factor $r_{h}/r_{J}$, with tidally-filling 
clusters evolving toward a less massive and more concentrated structure than 
underfilling ones. 
All the simulation starting with a Roche-lobe filling factor $r_{h}/r_{J}>0.15$ loose $>90\%$ 
of their initial mass and reach after 12 Gyr masses significantly smaller ($0.5<\Delta log(M/M_{\odot})<0.8$) than 
the mean present-day mass of GCs ($\langle log(M/M_{\odot})\rangle\sim 5.4$).

Summarizing, by analysing the set of simulations moving within the simplified 
point-mass galactic potential following circular orbits, it seems that only 
tidally-filling simulations with a significant degree of primordial mass 
segregation can reproduce the large fraction of SP found in Galactic GCs.
This configuration however lead to the emergence of SP with a degree of radial 
segregation which is incompatible with observations, unless strong tidal fields 
are considered.
This can be a big problem for the {\it cooling flow} scenario 
considering that $\langle r_{J5}\rangle$ varies by more than an order of 
magnitude among the Galactic GC system.

A solution is provided by the simulations moving in the axisymmetric 
potential of \citet{johnston1995} and following the eccentric orbit of NGC1851 
(N1851lm6.3w25rh34dr01 and N1851lm6.3w25rh17dr01ms).
These simulations, although starting in underfilling conditions, suffer strong 
mass loss during the peri-Galactic passages and the disk crossing. 
During the simulation, the half-mass radius is set at a reduced size with 
respect to a simulation moving on a circular orbit, being characterized by a 
shorter half-mass relaxation time. 
Consequently, both the fraction of SP stars and the $A_{2}^{+}$ turn out to be 
consistent with the observational values, in spite of the apparently large value of 
$\langle r_{J5}\rangle$. Moreover, in these simulations primordial mass 
segregation is not a necessary condition: also the simulation without 
primordial mass segregation (N1851lm6w25rh34dr01) can indeed reproduce the 
fraction of SP.

In Fig. \ref{agbmfall} the mass functions of FP and SP measured in the 
last snapshot of all simulations are shown. 
The mass functions of both populations are very similar, with 
slopes comprised in the range $-1.2<\alpha<-0.6$ and a maximum difference of 
$\Delta \alpha<0.15$ (i.e. with the FP with a slightly flatter MF than the SP), 
in agreement with what found by \citet{vesperini2018}.
As expected, because of the preferential loss of low-mass stars \citep{baumgardt2003}, simulations 
loosing a large fraction of their initial mass are those with the flattest 
mass function.

The anisotropy parameter $\beta\equiv 1-\sigma_{t}^{2}/\sigma_{r}^{2}$ (where 
$\sigma_{r}$ and $\sigma_{t}$ are the dispersions of the radial and tangential 
component of the velocity) is plotted as a function of the distance from the 
cluster centre in Fig.\ref{agbball} for all the 10 simulations.
Apart from small differences, in all the simulations the FP shows an anisotropy 
profile declining at large radii toward the regime of tangential anisotropy 
($\beta<0$), while the SP shows the opposite trend.
This result is almost independent on the initial degree of anisotropy of the FP 
(see simulations R14.4lm6w25rh36dr01ms and R14.4lm6w20rh36dr01ms\_anis).
This occurs because of the strong effect of tidal forces on the FP which favour 
the evaporation of stars on radial orbits. On the other hand, the SP is less 
affected by tides and preserves its original radial anisotropy.
The results presented here agree with those of \citet{henault2015}.
The simulations of \citet{tiongco2019} also predict a similar behaviour 
in the early stages of evolution while after many relaxation times they predict
isotropic profiles for both FP and SP. 
Given the extremely long relaxation time of the simulations presented 
here, the predictions of these works can be considered in agreement.

\subsection{Accretion onto proto-stellar disks scenario}
\label{disk_res_sec}

\begin{figure}
 \includegraphics[width=8.6cm]{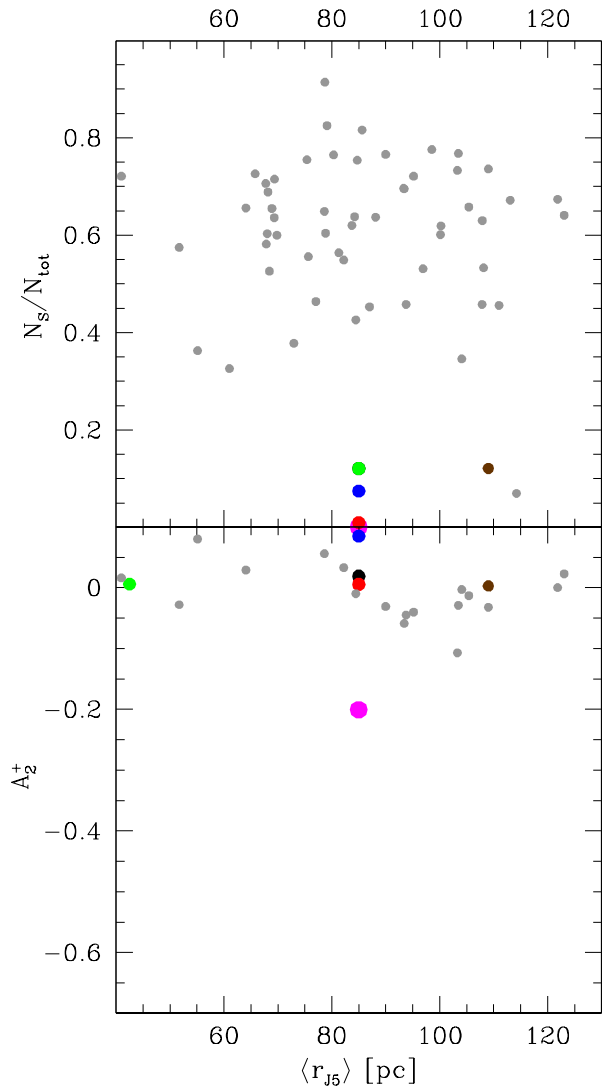}
 \caption{Same as Fig. \ref{agbres} but for the simulations run
 according to the {\it accretion onto proto-stellar disks} scenario.}
\label{mibres}
\end{figure}

\begin{figure}
 \includegraphics[width=8.6cm]{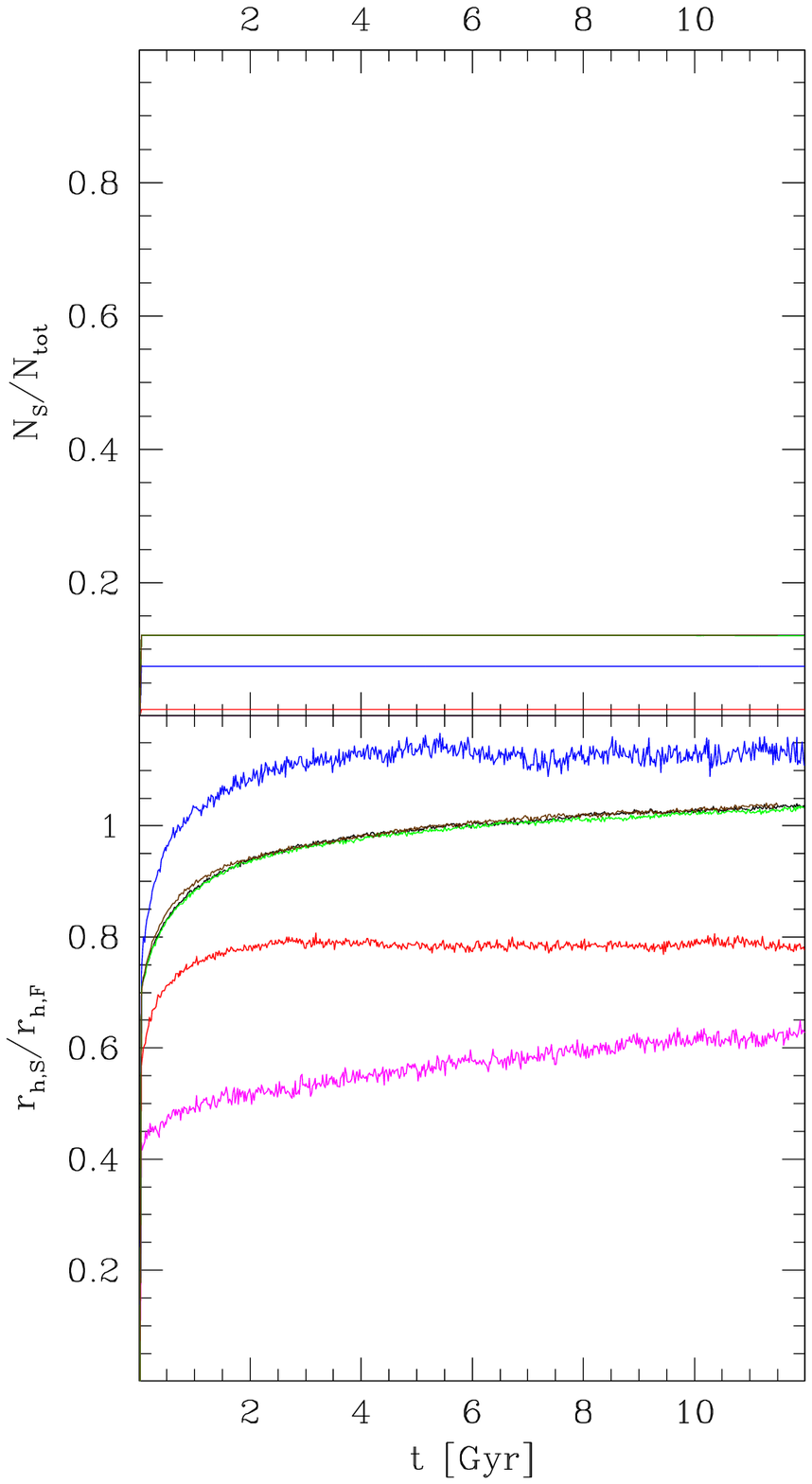}
 \caption{Same as Fig. \ref{agbev} but for the simulations run
 according to the {\it accretion onto proto-stellar disks} scenario.}
\label{mibev}
\end{figure}

\begin{figure}
 \includegraphics[width=8.6cm]{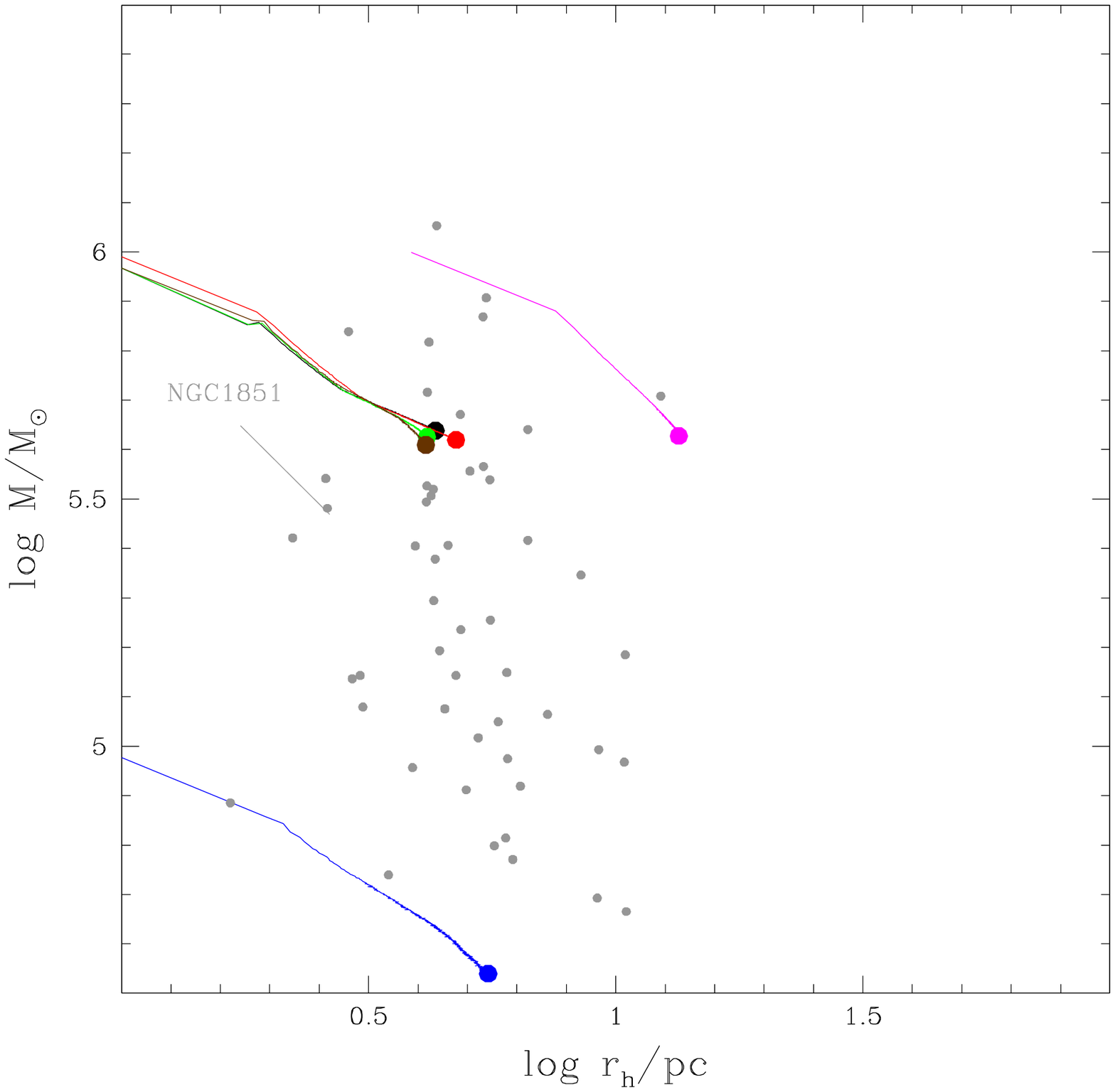}
 \caption{Same as Fig. \ref{agbmr} but for the simulations run
 according to the {\it accretion onto proto-stellar disks} scenario.}
\label{mibmr}
\end{figure}

\begin{figure*}
 \includegraphics[width=\textwidth]{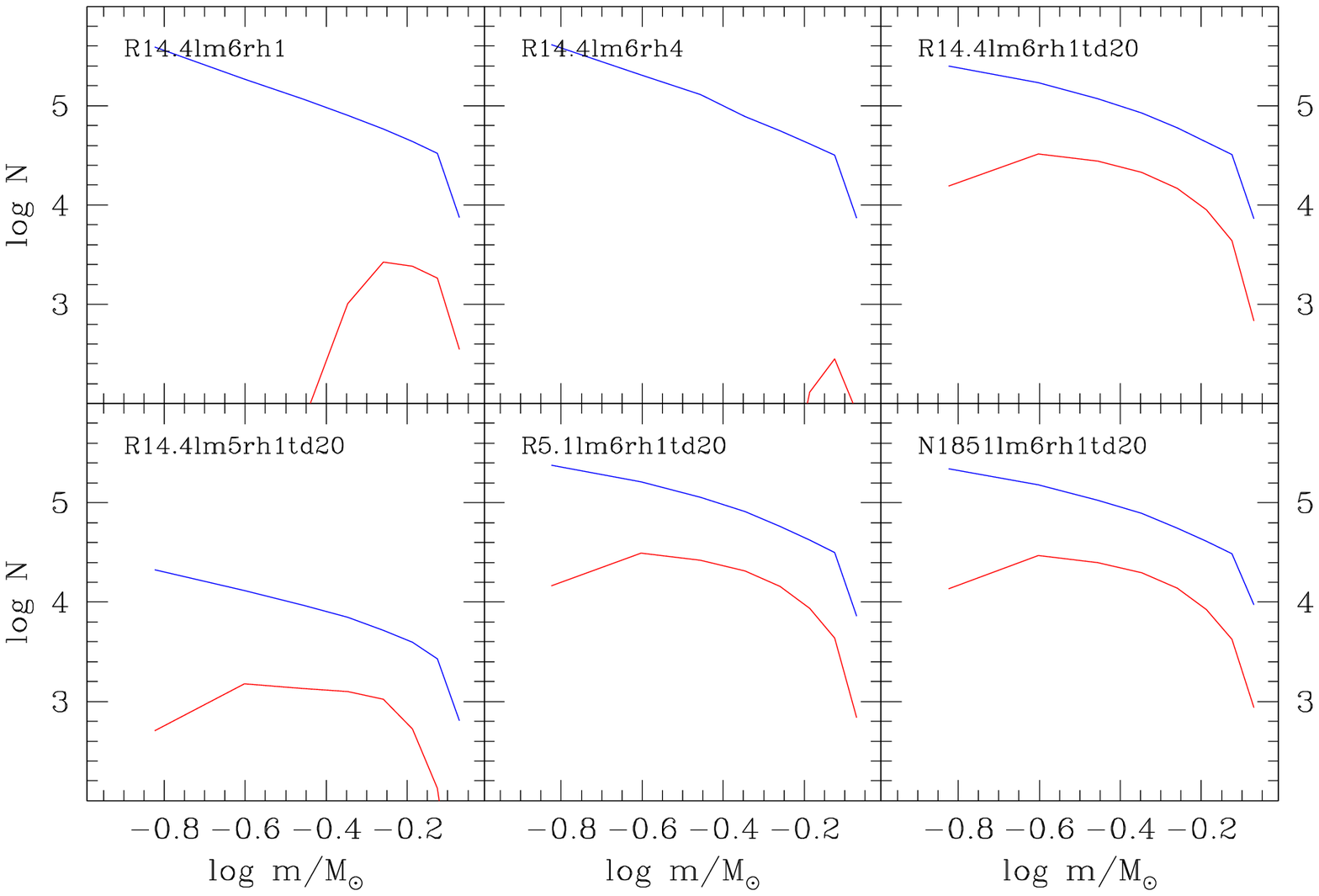}
 \caption{Same as Fig. \ref{agbmfall} but for the simulations run
 according to the {\it accretion onto proto-stellar disks} scenario.}
\label{mibmfall}
\end{figure*}

\begin{figure*}
 \includegraphics[width=\textwidth]{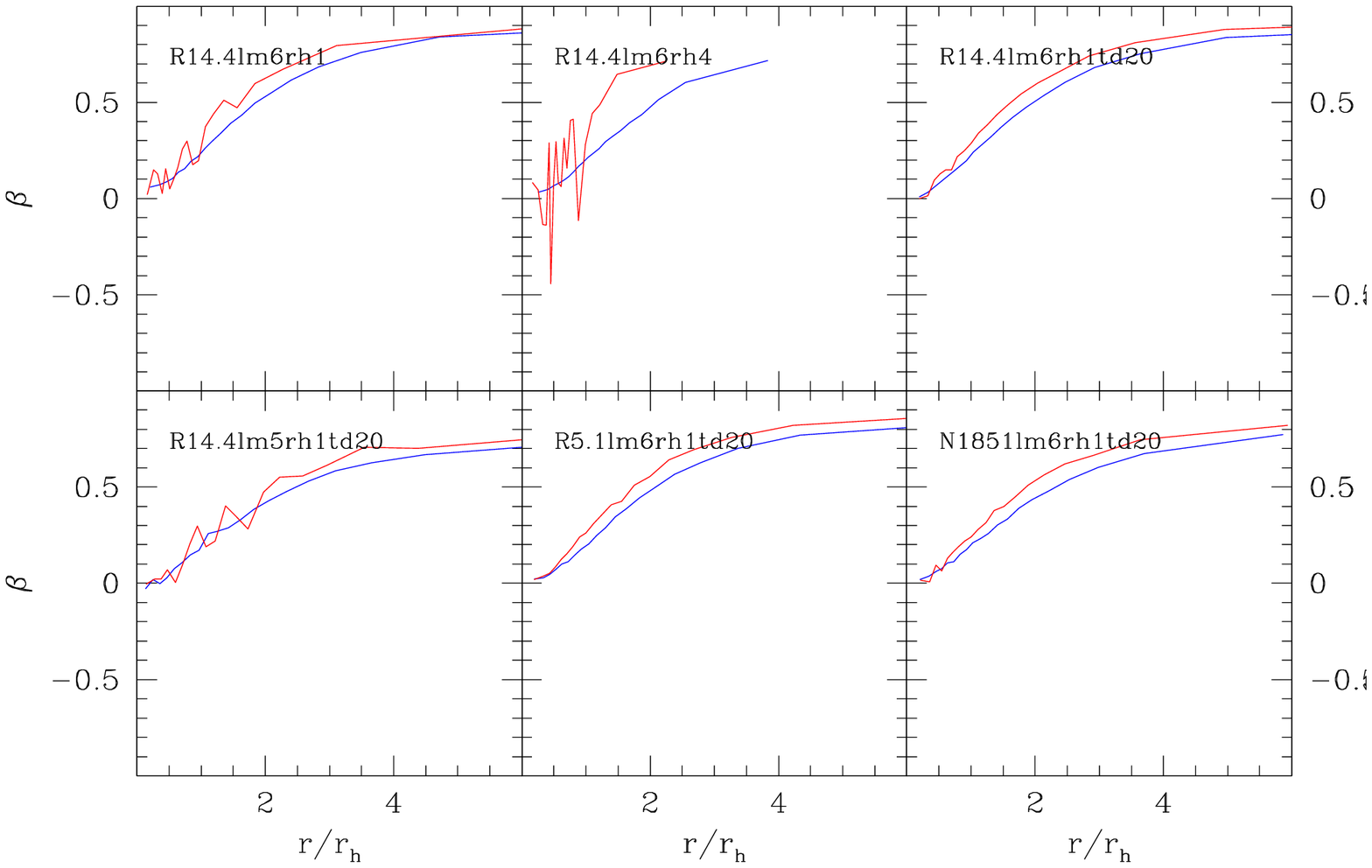}
 \caption{Same as Fig. \ref{agbball} but for the simulations run
 according to the {\it accretion onto proto-stellar disks} scenario.}
\label{mibball}
\end{figure*}

In Fig. \ref{mibres} the
fraction of SP and the $A_{2}^{+}$ value measured un the last snapshot of the 6
simulations run according to the {\it accretion onto proto-stellar disks} scenario.
All the 6 simulations lead to very small SP fraction ($N_{S}/N_{tot}<0.13$).
This is not surprising given the extremely small initial fraction of SG (see 
Sect. \ref{disk_setup_sec}). In this scenario all the simulations start in 
underfilling conditions both to form a significant SP and to reproduce the 
present-day half-mass radius of GCs.
Moreover, the SP is only mildly more concentrated than FP at the beginning of 
these simulations.
So, both FP and SP are free to expand and interact during the subsequent 
evolution without a preferential loss of FP stars, thus leaving the 
fraction of SP stars almost unchanged. This is apparent in the top panel 
of Fig. \ref{mibev} where the evolution of the SP fraction is shown for 
all the 6 simulations.

Regarding the radial distribution of SP, all the simulations with an 
increased disk lifetime have SP completely mixed with FP.
As shown in the bottom panel of Fig. \ref{mibev}, in compact and
underfilling conditions, two-body relaxation is efficient in mixing the two 
populations at early epochs, while the dynamical evolution at later times 
($t>2~Gyr$; after the stellar evolution-driven expansion) only 
marginally affects the radial segregation of SP.
In this respect, simulations with extended disk lifetimes host a SP 
containing a larger fraction of low-mass stars (see Fig. \ref{mibprof}) and 
a more extended radial distribution with 
respect to simulations with the disk lifetimes predicted by the LX16 model. 

In Fig. \ref{mibmr} the evolution of the 6 considered simulations in the 
mass vs. half-mass radius plane is shown. Most simulations show the same 
behaviour with an expansion (driven by stellar evolution) and a modest mass loss 
($\Delta log (M/M_{\odot})\sim 0.3$; almost entirely addressed to stellar 
evolution mass loss). In this case, simulations with an initial mass of 
$M=10^{6}~M_{\odot}$ and half-mass radius $r_{h}=1~pc$ well reproduce after 
12 Gyr the average present-day mass and half-mass radius of Galactic GCs.

Note also that in this scenario the simulations run in a simplified potential 
(R14.4lm6rh1td20 and R5.1lm6rh1td20) and that run in the 
axisymmetric potential (N1851lm6rh1td20) produce very similar results.
Indeed, in all these simulations, the entire cluster is always 
contained in a small volume well inside the Jacobi radius.

In Fig. \ref{mibmfall} the mass functions of the two populations at the 
end of the 6 simulations are compared. The strong mass function difference observed at the 
beginning of the simulation (Sect. \ref{disk_setup_sec}) remains 
apparent also after 12 Gyr in all the considered simulations. 
This is not surprising given the relatively 
small amount of mass lost by both populations. 
In particular, while the FP has a mass function consistent with a power-law, 
the mass function of the SP is significantly depleted of low-mass 
($M<0.3~M_{\odot}$) stars. 

The anisotropy profiles measured in the last snapshot of the 6 simulations 
are shown in Fig. \ref{mibball}. In all simulations, both populations 
show a radially anisotropic profile. 
This is a consequence of the many 
close encounters occurring in the dense cluster centres which eject stars 
of both populations
on radial orbits at large distances \citep{lyndenbell1968,spitzer1975}.
The behaviour of the anisotropy profiles shown in Fig. \ref{mibball} 
differs from that predicted by the simulations of \citet{henault2015}.
In the simulations of these authors indeed the FP is isotropic at the end of 
the evolution, similarly to what happens in the {\it cooling flow} scenario, 
while the SP is radially anisotropic at large radii.
The difference between the two works likely resides in the different initial structure of the simulations. 
The simulations of \citet{henault2015} are indeed $\sim$1000 time 
less dense than the simulations run here, so that the number of close collisions 
occurring in their centres is much lower.

\section{Summary}
\label{concl_sec}

I presented the results of a large set of Monte Carlo simulations of star clusters
run adopting the initial conditions predicted by the two main scenarios proposed 
so far for the formation of multiple populations.
These simulations, containing $1.5\div 3\cdot 10^{6}$ particles, are among the largest 
simulations ever run and allow for the first time to reach final masses similar or 
only a factor $\sim 5$ smaller than those of real GCs.

For the {\it accretion onto proto-stellar disks} scenario, the accretion rate 
of individual disks has been modeled for the first time, accounting for the 
evolution of the size of the proto-stellar disk.
Because of the short disk-lifetimes, only a negligible fraction of polluted 
material can be accreted onto the proto-stellar disks of low-mass stars.
Even adopting a mass-independent disk lifetime of 20 Myr, the SP do not 
exceed 10\% of the whole stellar content.
Moreover, the continuous distribution of orbital energies leads to a continuous 
distribution of the time spent by stars within the polluted gas cloud, thus 
avoiding the formation of discrete populations.
Because of the small disk radii at low masses and the quick disappearance of 
the convective proto-stellar core, the IMF of SP is depleted of low- ($M<0.3~M_{\odot}$) 
and high-mass ($M>1.5~M_{\odot}$) stars.
The subsequent dynamical evolution does not significantly change any of the 
above initial differences. Indeed, the initial density required to constitute 
a significant amount of polluted gas implies an initially concentrated structure 
($r_{h}(0)=1~pc$), so that the cluster remains extremely resistant to tidal effects and 
retains a large fraction of stars of both populations. 
The final fraction of SP turns out to be therefore inconsistent with those observed 
today in Galactic GCs \citep{milone2017}.
Several works foresaw the difficulty of this scenario in producing discrete populations 
\citep[see e.g.][]{renzini2008} and in maintaining an efficient convective core 
for a significant amount of time \citep{dantona2014}, although this is the first time 
these problems are quantified. These issues adds to the already known problems of this 
scenario in reproducing the Li abundance of SP \citep{dantona2014} and the stability 
of disks in a strong-accretion regime \citep{wijnen2016}. 

The simulations run under the {\it cooling flow} scenario, given the large 
number of involved free parameters, are instead
able to reproduce at the end of their evolution the present-day general 
properties of FP/SP (in terms of their relative fraction and radial segregation) 
reaching a global structure comparable (i.e. a similar half-mass radius and a 
slightly lower mass) to those of Galactic GCs.
The imperative conditions to provide a good agreement are however that the cluster {\it i)} 
form in a tidally-filling configuration ($r_{h}/r_{J}>0.15$; although this condition depends on the initial SP/FP mass ratio) and {\it ii)} maintains 
a compact structure during its evolution.
The first condition ensures a large fraction of SP stars, while the second 
allows the spatial mixing of the two populations. 
Indeed, the larger is the filling-factor, the more efficient is the loss of FP 
determining the final proportion of SP/FP.
On the other hand, the smaller is the cluster size the shorter is 
the relaxation time, thus increasing the efficiency of two-body relaxation.
The lack of a decreasing trend between the relative fraction of SP with the Galactocentric 
distance implies that the above conditions must be fulfilled at all 
Galactocentric distances. This can represent a big problem for the 
{\it cooling flow} scenario.
Indeed, in a logarithmic potential \citep[like the one of the Milky Way halo;][]{johnston1995},
the Jacobi radius grows as $r_{J}\propto R_{g}^{2/3}$. So, to keep the 
filling-factor fixed, the initial half-mass radius of the cluster should grow accordingly.
It is therefore impossible to satisfy both the above conditions across the large range 
of Galactocentric distances \citep[$0.5<R_{g}/kpc<125$][2010 edition]{harris1996} covered by Galactic GCs.
In particular, at large Galactocentric distances GCs the Galactic tidal field 
is so weak that the cluster spend all its life with a large half-mass radius 
constituting a non-collisional environments where the SP would maintain a strong 
radial segregation (reaching $r_{h,S}/r_{h,F}\sim 0.2$), much stronger than what observed 
among Galactic GCs. This problem would be further amplified if smaller fractions of SP/FP would be assumed \citep[as proposed by ][]{cabrera2015}.
Moreover, it is difficult to explain how star formation can ignite in the extremely 
low-density environment required by this scenario in GCs in the outer Galactic halo. 
A solution to this problem can be linked to the orbits distribution of Galactic GCs.
The two simulations run in the axisymmetric potential of \citet{johnston1995} and following 
the eccentric orbit of NGC1851 both conclude their evolutions with a SP with 
a reasonable fraction and radial distribution in spite of their large present-day 
Galactocentric distance. In these cases, disk- and bulge-shocks contribute to trigger 
the mass loss of FP and to keep the cluster in a compact configuration, even if the cluster 
starts in an underfilling condition.
Of course, the shape of the Galactic potential and the orbital eccentricity cannot erase
the weakening of the tidal strength at large Gactocentric distances. 
However, it is known that GCs orbits become radially anisotropic in the outer halo 
\citep{vasiliev2019}, and this can contribute to maintain a significant tidal 
truncation also for GCs at large Galactocentric distances.
Unfortunately, only a single eccentric orbit has been considered here, so that it 
is not clear if such an effect can solve the above problem for the GCs along 
the entire range of Galactocentric distances.
Further surveys of simulations, tailored to reproduce the observed SP/FP population ratios 
and following the actual orbit of a significant number of GCs are necessary to clarify this issue.

It is interesting to note that the two considered scenarios foresee
opposite predictions for the mass functions and the anisotropy profiles of FP and SP. 
The determination of these quantities can therefore be good tools to 
distinguish between these two scenarios.
In particular, while in the {\it cooling flow} scenario no significant 
differences are expected in the mass function of FP and SP, in the {\it accretion onto 
proto-stellar disks} scenario the SP should show a significant depletion of low-mass 
($M<0.3~M_{\odot}$) masses.  
Unfortunately, the only available study of this kind \citep{milone2012} is focused on the 
very central region of the cluster where the effect of mass segregation produces 
strong modifications to the global mass function.
Similarly, while in the {\it cooling flow} scenario the FP is characterized by a significantly 
smaller degree of radial anisotropy than the SP, in the {\it accretion onto 
proto-stellar disks} scenario the two populations are expected to share the same anisotropic profile.
However, observational studies provide conflicting results, 
with some clusters hosting stellar populations with similar anisotropy profiles and 
others with significant differences \citep{cordoni2020}.

In the present paper, I do not compare the predicted behaviour of other kinematic parameters
of FP and SP like e.g. binary and remnant fraction. 
While these quantities contain important information, their present-day appearance 
strongly depend on their uncertain initial conditions.
Future work will be addressed to investigate the evolution of these quantities spanning 
a wide range in their initial conditions.

\section*{Acknowledgments}

I warmly thank Enrico Vesperini for useful discussions. I also thank the anonymous referee for his/her helpful comments and suggestions.

\section*{Data availability}

The data underlying this article will be shared on reasonable request to the corresponding author.

\begin{landscape}
%\centering
\begin{table}
  \caption{Initial and final properties of the performed simulations.}
  \begin{tabular}{@{}lcccccccc@{\hskip 1cm}cccccl@{}}
 \hline
                           & \multicolumn{8}{c}{Initial} & \multicolumn{4}{c}{Final} & &\\
 name                      & $R_{g}$ & Mass        & $W_{0}$ & $M_{F}$            & $r_{h}$ & $M_{S}/M_{tot}$ & $r_{h,S}/r_{h,F}$ & $r_{h}/r_{J}$ & M                  & $r_{h}$ & $N_{S}/N_{tot}$ & $A_{2}^{+}$ & colour & $t_{disk}$\\
                           &         & segregation &         &                    &         &                 &                   &               &                    &         &                  &            &        & /Note\\
                           & kpc     &             &         & $10^{5}~M_{\odot}$ &   pc    &                 &                   &               & $10^{5}~M_{\odot}$ &   pc    &                  &            &        & Myr \\
  \hline
\multicolumn{15}{l}{{\it Cooling flow} scenario}\\
\hline
 R14.4lm6w5rh36dr01         & 14.4  & no & 5 & 10 & 36 & 0.091 & 0.10 & 0.200 & 2.21 & 22.68 & 0.235 & -0.546 & red &\\
 R14.4lm6w3rh36dr01         & 14.4  & no & 3 & 10 & 36 & 0.091 & 0.10 & 0.200 & 1.57 & 21.66 & 0.329 & -0.585 & magenta &\\
 R14.4lm6w25rh36dr01ms      & 14.4  & yes & 25 & 10 & 36 & 0.091 & 0.10 & 0.200 & 0.57 & 9.08 & 0.812 & -0.336 & black &\\
 R14.4lm6w25rh18dr01ms      & 14.4  & yes & 25 & 10 & 18 & 0.091 & 0.10 & 0.100 & 4.94 & 32.30 & 0.105 & -0.093 & orange &\\
 R5.1lm6w25rh18dr01ms     & 5.1 & yes & 25 & 10 & 18 & 0.091 & 0.10 & 0.200 & 0.43 & 4.97 & 0.818 & -0.672 & green &\\
 R14.4lm6w25rh36dr005ms     & 14.4  & yes & 25 & 10 & 36 & 0.091 & 0.05 & 0.200 & 0.51 & 7.74 & 0.825 & -0.365 & cyan &\\
 R14.4lm6w25rh36dr02ms      & 14.4  & yes & 25 & 10 & 36 & 0.091 & 0.20 & 0.200 & 0.47 & 12.91 & 0.845 & -0.222 & blue &\\
 R14.4lm6w20rh36dr01ms\_anis & 14.4  & yes & 20 & 10 & 36 & 0.091 & 0.10 & 0.200 & 1.06 & 18.88 & 0.485 & -0.525 & indigo & anis FP/isot SP\\
 N1851lm6.3w5rh34dr01	   & NGC1851-like & no & 5 & 20 & 34 & 0.091 & 0.10 & 0.094 & 0.75 & 5.73 & 0.710 & -0.111 & brown &\\
 N1851lm6.3w25rh17dr01ms   & NGC1851-like & yes & 25 & 20 & 17 & 0.091 & 0.10 & 0.047 & 0.79 & 4.06 & 0.824 & -0.072 & brown-dashed &\\
 \hline
\multicolumn{15}{l}{{\it Accretion onto proto-stellar disks} scenario}\\
\hline
 R14.4lm6rh1        & 14.4  & yes & 25 & 10   & 1 & 0.012 & 0.57 & 0.005 & 4.17 & 4.75  & 0.010 &  0.001 & red & LX16\\
 R14.4lm6rh4        & 14.4  & yes & 25 & 10   & 4 & 0.002 & 0.43 & 0.022 & 4.25 & 13.42 & 0.001 & -0.188 & magenta & LX16\\
 R14.4lm6rh1td20    & 14.4  & yes & 25 & 10   & 1 & 0.082 & 0.70 & 0.005 & 4.35 & 4.32  & 0.121 &  0.019 & black & 20\\
 R14.4lm5rh1td20    & 14.4  & yes & 25 & 1    & 1 & 0.052 & 0.73 & 0.005 & 0.35 & 5.52  & 0.075 &  0.114 & blue & 20\\
 R5.1lm6rh1td20   & 5.1 & yes & 25 & 10   & 1 & 0.082 & 0.70 & 0.005 & 4.23 & 4.15  & 0.121 &  0.014 & green & 20\\
 N1851lm6rh1td20   & NGC1851-like & yes & 25 & 10 & 1 & 0.082 & 0.70 & 0.003 & 4.07 & 4.13 & 0.121 & 0.012 & brown & 20\\
\hline
\end{tabular}
 \label{table1_tab}
\end{table}
\end{landscape}

\label{lastpage}
\end{document}